# Belief Propagation Decoding of Polar Codes on Permuted Factor Graphs

Ahmed Elkelesh, Moustafa Ebada, Sebastian Cammerer and Stephan ten Brink
Institute of Telecommunications, Pfaffenwaldring 47, University of Stuttgart, 70569 Stuttgart, Germany
{elkelesh,ebada,cammerer,tenbrink}@inue.uni-stuttgart.de

*Abstract*—We show that the performance of iterative belief propagation (BP) decoding of polar codes can be enhanced by decoding over different carefully chosen factor graph realizations. With a genie-aided stopping condition, it can achieve the successive cancellation list (SCL) decoding performance which has already been shown to achieve the maximum likelihood (ML) bound provided that the list size is sufficiently large. The proposed decoder is based on different realizations of the polar code factor graph with randomly permuted stages during decoding. Additionally, a different way of visualizing the polar code factor graph is presented, facilitating the analysis of the underlying factor graph and the comparison of different graph permutations. In our proposed decoder, a high rate Cyclic Redundancy Check (CRC) code is concatenated with a polar code and used as an iteration stopping criterion (i.e., genie) to even outperform the SCL decoder of the plain polar code (without the CRC-aid). Although our permuted factor graph-based decoder does not outperform the SCL-CRC decoder, it achieves, to the best of our knowledge, the best performance of all iterative polar decoders presented thus far.

## I. INTRODUCTION

Recently, polar codes [1] have been considered as channel codes for the upcoming *5th generation mobile communication standard (5G)* as part of the control channel [2] and thus, decoding of polar codes has more and more become a practical implementation challenge. As successive cancellation (SC) decoding [1] for finite length polar codes is sub-optimal, successive cancellation list (SCL) decoding [3] is applied, at the cost of increased complexity due to the list decoding nature. It was shown in [3] that this decoder can approach the maximum likelihood (ML) bound for a sufficiently large list size. Furthermore, an additional high rate Cyclic Redundancy Check (CRC) code can be easily used under SCL decoding in order to enhance the code itself via increasing its minimum distance. This combination renders polar codes into a powerful coding scheme. However, SCL suffers from both, high-complexity and an inherently serial decoding nature.

In this work, we focus on iterative decoding of polar codes based on the message passing algorithm over the encoding graph of polar codes [4]. In contrast to SCL decoding, the belief propagation (BP) algorithm can be easily parallelized [5]. Additionally, this approach inherently enables soft-in/soft-out decoding and allows joint iterative detection and decoding loops. Although the BP algorithm can outperform SC decoding, it is yet not competitive to SCL decoding and, thus, not

This work has been supported by DFG, Germany, under grant BR 3205/5-1.

attractive for many applications. In [6], [7], it has been shown that the finite length polar codes under BP decoding can be enhanced when the *semi-polarized* channels are additionally protected by check nodes, or an augmented (shorter) polar code [8]. Unfortunately, these approaches are still of an inferior performance when compared to SCL decoding. Besides, they require an adjusted code structure and are thus not compatible to the expected standardized polar code (i.e., polar code concatenated with CRC code). The proposed algorithm in this paper enhances the decoder without any modifications needed at the encoder. The proposed algorithm works well with any polar code concatenated with a CRC code.

For a polar code of length $N$, the stages of the encoding graph can be permuted [9] leading to $(\log_2 N)!$ different graphs with the same encoding behavior. As a result, an almost infinite amount of different *decoder* permuted realizations exist even for moderate length polar codes (e.g., $10! > 4 \times 10^8$ permutations for an $N = 1024$ polar code). The individual factor graph permutation performance per codeword and noise realization is different due to the different order of processing in the decoding graph. In other words, whenever decoding does not succeed on a specific graph, another graph permutation can be used until reaching a specific stopping condition. Decoding on different factor graph permutations in parallel and combining all obtained decoding results was first mentioned in [9], but has not been investigated further.

This idea is somewhat akin to the idea of redundant Tanner graphs for decoding of LDPC codes [10]. Similar ideas have been successfully applied to high density parity check (HDPC) codes, e.g., Reed-Solomon (RS) codes in [11]. Recently, the combination of random redundant iterative decoders [12] and machine learning techniques for Bose-Chaudhuri-Hocquenghem (BCH) codes has been shown in [13].

In the following, we show that genie-aided iterative decoding can approach SCL performance, and thus also ML performance, when decoding over multiple factor graph permuted realizations. Although the decoding complexity (and maximum latency) seems to increase drastically in our basic implementation, it is shown in Subsection IV-C that the increase in complexity is not as dramatic in the low error-rate region. However, the main objective of this work is to show the potential of iterative decoding and the effects of multiple decoding graph permutations; therefore, we keep the complexity reduction open for future work.

## II. POLAR CODES AND ITERATIVE DECODING

In this section, we briefly mention fundamental concepts of polar codes, besides introducing the notations used throughout this work. We also briefly discuss the BP decoder of polar codes and review some of the recent work pursued on it.

### A. Polar Codes

Polar codes introduced in [1] are constructed based on a $2 \times 2$ kernel (i.e., Arıkan's kernel $\mathbf{G}_2 = \begin{bmatrix} 1 & 0 \\ 1 & 1 \end{bmatrix}$ is the most commonly used). After recursively acquiring the $n^{th}-$kronecker power of the kernel $\mathbf{G}_2$, $N$ synthesized bit channels are obtained, where $N = 2^n$. The term "polarized channels" indicates that one portion of the synthesized channels is purely noisy, and thus, would be frozen (i.e., cannot be used for useful data transmission and would be set to known values, e.g., 0 in this work), and another portion would be purely noiseless which will be the information bit channels, i.e., used for data transmission. However, the phenomenon of channel polarization requires sufficiently large $N$, where the channels converge to either purely noisy or purely noiseless channels.

The selection of *good* and *bad* bit channels out of the $N$ synthesized channels is called *code construction*, where the information set $\mathbb{A}$ is the set of indices denoting the information bit channels. Several code construction algorithms exist, with different bit channel "quality" criteria, e.g., [14][15][16]. Throughout this paper, we use the polar code construction based on Arıkan's Bhattacharyya bounds of bit channels [1], however, any other polar code construction algorithm could be used straightforwardly.

A polar code of length $N$ with $k$ information bits and code rate $R = \frac{k}{N}$ is denoted by $\mathscr{P}(N,k)$. Encoding requires the computation of the $N \times N$ generator matrix $\mathbf{G}_N$ by computing the kronecker product $\mathbf{G}_2^{\otimes n}$. A vector $\mathbf{u}$ of length $N$ is constructed containing $k$ information bits placed in the $\mathbb{A}$ indices and zeros in the remaining indices $\bar{\mathbb{A}}$. The $N$ coded bits $\mathbf{x}$ are calculated as follows $\mathbf{x} = \mathbf{u} \cdot \mathbf{G}_N$.

For the family of polar codes, there exist two main decoding schemes: SC decoding (and its variants, e.g., SCL) and BP decoding. Although polar codes were theoretically proven to achieve the symmetric channel capacity of a Binary Input Discrete Memoryless Channel (BI-DMC) under SC decoding assuming an infinite length code, finite length polar codes show a degraded performance because of the incomplete channel polarization phenomenon [6]. An alternative iterative decoding algorithm based on the idea of message passing over the encoding graph was introduced in [4], and was shown to outperform the SC decoding for finite length polar codes.

### B. Belief Propagation Decoding

The flooding BP decoding of polar codes is a message passing algorithm in which the information bits are retrieved through iterations conducted on the factor graph corresponding to the polar code generator matrix $\mathbf{G}_N$. As depicted in Fig. 1a, the polar code factor graph consists of $n = \log_2 N$ stages. In the following, all messages are assumed to be log-likelihood ratios (LLR) and are defined as

$$LLR(x) = \log \frac{P(x=0|y)}{P(x=1|y)}.$$

*1) Types of LLR messages:* Two types of messages are involved, left-to-right messages (**R**-messages) and right-to-left messages (**L**-messages). The **R**-messages at stage 1 represent the a priori information available to the decoder and, thus, are either zero or infinity for non-frozen and frozen bits, respectively. The **L**-messages at stage $n+1$ carry the LLR channel output $\mathbf{L}_{ch}$. All other messages are initialized with zero (i.e., no initial information).

*2) Types of LLR updates:* Each single iteration is composed of one left-to-right message propagation, updating the LLR values of the **R**-messages and one right-to-left message propagation, updating the LLR values of the **L**-messages.

*3) Factor graph and processing element (PE):* The polar factor graph (see Fig. 1) consists of $\frac{N}{2} \cdot \log_2(N)$ PEs. The **L**- and **R**-messages are updated in each PE (shown in [17, Fig. 2]) as follows:

$$R_{\text{out},1} = f(R_{\text{in},1}, L_{\text{in},2} + R_{\text{in},2})$$
$$R_{\text{out},2} = f(R_{\text{in},1}, L_{\text{in},1}) + R_{\text{in},2}$$
$$L_{\text{out},1} = f(L_{\text{in},1}, L_{\text{in},2} + R_{\text{in},2})$$
$$L_{\text{out},2} = f(R_{\text{in},1}, L_{\text{in},1}) + L_{\text{in},2}$$

where $f(L_1, L_2) = L_1 \boxplus L_2$ is commonly referred to as *boxplus* operator, which can be expressed as

$$f(x,y) = x \boxplus y = \log \frac{1 + e^{x+y}}{e^x + e^y}.$$

*4) Decoding termination and stopping conditions:* Traditionally, the conventional BP decoder terminates when it reaches a predefined maximum number of BP iterations $N_{it,max}$. The LLRs of the estimated message $\hat{\mathbf{u}}$ and the estimated transmitted codeword $\hat{\mathbf{x}}$ are computed according to

$$L(\hat{u}_i) = L_{1,i} + R_{1,i}$$
$$L(\hat{x}_i) = L_{n+1,i} + R_{n+1,i}$$

However, early stopping conditions introduced in [18] are used to speed up the decoding process by terminating the decoding process if a certain stopping condition is met. One of the conditions proposed is **G**-matrix based, where $\hat{\mathbf{u}}$ is said to be a valid estimate of $\mathbf{u}$ if $\hat{\mathbf{x}} = \hat{\mathbf{u}} \cdot \mathbf{G}$ is fulfilled. Throughout this work, this stopping condition is called *"practical stopping condition"*. Note that this multiplication is of high complexity and, thus, can be avoided by encoding over the polar encoding circuit which has a complexity of $\mathscr{O}(N \cdot \log N)$. One can obviously infer that if the BP decoder is terminated using the condition $\hat{\mathbf{u}} = \mathbf{u}$, this would act like a lower bound on the BP decoder performance (continue iterating till reaching the correct transmitted information bits, i.e., perfect knowledge-based decoding). Throughout this work, this stopping condition is called *"perfect knowledge-based stopping condition"*, since it is more of a bound rather than representing a real decoding

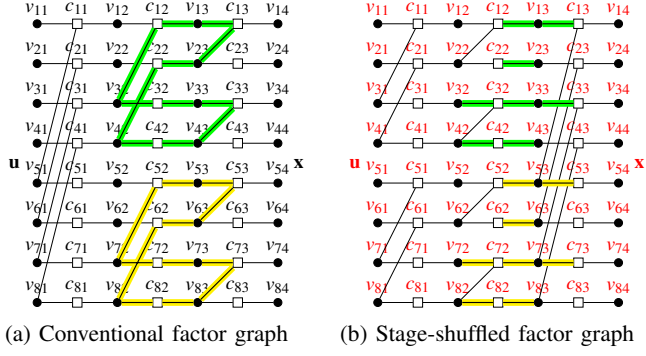

(a) Conventional factor graph  (b) Stage-shuffled factor graph

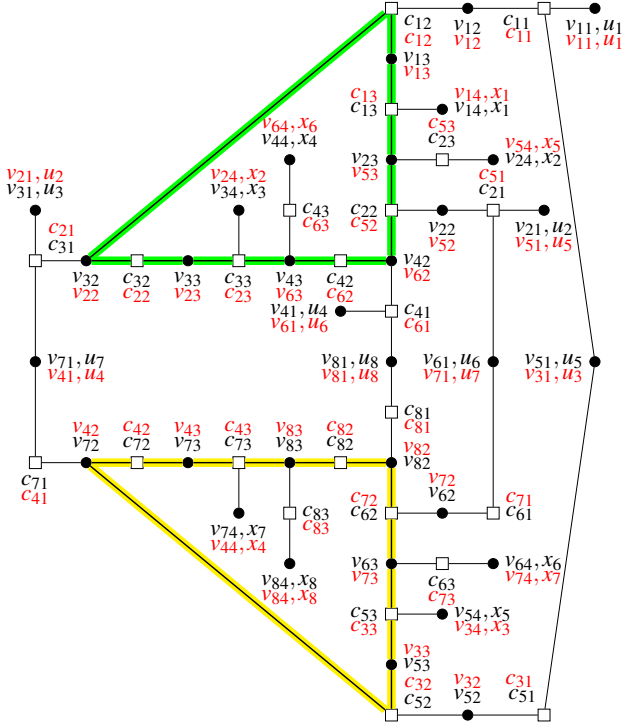

(c) Unfolded factor graph labeled according to (a) and (b)

Fig. 1: Conventional and permuted factor graph of $\mathscr{P}(8,k)$.

behavior. A further stopping (and detection) condition was introduced in [19], where a high rate CRC code is used as an outer code on the information bit vector to overcome the situation of undetected errors (where the estimate $\hat{\mathbf{x}}$ is a valid codeword but $\hat{\mathbf{x}} \neq \mathbf{x}$). Thus, stop the BP iterations when the CRC on the information bits is satisfied.

### III. DECODING ON PERMUTED FACTOR GRAPHS

As exemplary shown in Fig. 1 for two permutations, there are $(\log_2 N)!$ different permutations of the polar code factor graph based on the generator matrix $\mathbf{G}$. The BP decoding algorithm can be performed over any of such permutations [9]. The leftmost stage (i.e., stage 1) of all of the different permutations of the polar factor graph represents the vector $\mathbf{u}$ containing the frozen known bits and the non-frozen information bits. The rightmost stage (i.e., stage $n+1$) of all of the different permutations of the polar factor graph represents the codeword $\mathbf{x}$ or its corresponding LLRs.

Polar decoding on permuted factor graphs was referred to as "BP decoding on an *overcomplete representation* of a factor graph" in [9], or simply "multi-trellis" BP decoding. It was used in [17] to overcome error floors due to inadequate LLR-clipping values. This decoding algorithm (Algorithm 1) works as follows. BP decoding iterations are performed over a random permutation of the polar code factor graph shown in Fig. 1a until a certain early stopping condition (Algorithm 2) [18] is fulfilled, or until a predefined maximum number of BP iterations per trellis $N_{it,max}$ is reached. If decoding on one factor graph permutation fails (i.e., stopping condition is never satisfied and a maximum number of iterations is reached), the information from the channel $\mathbf{L}_{ch}$ and the a priori information to the decoder (i.e., frozen and non-frozen bit positions) are passed on to a new factor graph permutation (e.g., Fig. 1b). When successively reaching a predefined maximum number of factor graph permutations $q_{max}$, decoding terminates. This can be viewed as a multi-stage decoding process in which a new stage is invoked if the previous stage(s) failed to converge.

---

**Algorithm 1** Multi-trellis BP decoder

---

**Input:**

$\mathbf{L}_{ch}$, ▷ LLR channel output
$\mathbb{A}$, ▷ information set
$q_{max}$, ▷ max. no. of factor graphs
$N_{it,max}$, ▷ max. no. of iterations per factor graph
$\mathbf{u}$, ▷ transmitted information bits
$stopID$. ▷ stopping criterion

**Output:**

$\hat{\mathbf{u}}$. ▷ estimated information bits

1: $N \leftarrow length(\mathbf{L}_{ch})$
2: $n \leftarrow \log_2 N$
3: $iq \leftarrow 1$
4: **for** $iq \leq q_{max}$ **do**
5: $\quad (\mathbf{L}, \mathbf{R}) \leftarrow \textbf{initializeLandR}(\mathbf{L}_{ch}, \mathbb{A})$
6: $\quad schedule \leftarrow \textbf{permute}(\{1,...,n\})$
7: $\quad iI \leftarrow 1$
8: $\quad$ **for** $iI \leq N_{it,max}$ **do**
9: $\quad\quad (\mathbf{L}, \mathbf{R}) \leftarrow \textbf{OneBPiteration}(N, n, \mathbf{L}, \mathbf{R}, schedule)$
10: $\quad\quad$ **if** checkStopCondition$(\mathbf{L}, \mathbf{R}, \mathbf{u}, stopID)$ **then**
11: $\quad\quad\quad \hat{\mathbf{u}} = \text{LLR2bit}(\mathbf{L}(1,:) + \mathbf{R}(1,:))$
12: $\quad\quad\quad$ **return** $\hat{\mathbf{u}}$
13: $\quad\quad$ **end if**
14: $\quad\quad iI \leftarrow iI + 1$
15: $\quad$ **end for**
16: $\quad iq \leftarrow iq + 1$
17: **end for**
18: $\hat{\mathbf{u}} = \text{LLR2bit}(\mathbf{L}(1,:) + \mathbf{R}(1,:))$
$\quad\quad\quad\quad\quad$ ▷ no stop. cond. satisfied
19: **return** $\hat{\mathbf{u}}$

---

The soft values in the intermediate stages of a specific factor graph are cleared in subsequent decoding iterations on a permuted factor graph realization. This is because the information that $\mathbf{L}$- and $\mathbf{R}$-messages at certain stages hold is different for each factor graph realization. This can be inferred from Fig. 1c. Some soft messages are shared between different realizations of the factor graph, and thus, these messages can be re-used (i.e, not dismissed). Some soft messages in a certain factor graph represents certain messages which are not explicitly seen in another factor graph (i.e., should be either dismissed or somehow translated according to the new factor graph permutation). For simplicity we clear all internal messages after each factor graph permutation in this work.

**Algorithm 2** checkStopCondition

**Input:**
  **L**,          ▷ L-matrix of BP factor graph
  **R**,          ▷ R-matrix of BP factor graph
  **u**,          ▷ transmitted information bits
  *stopID*.       ▷ stopping criterion

**Output:**
  *isSatisfied*.  ▷ indicate if stop. condition is satisfied

1: $\hat{\mathbf{u}} \leftarrow \text{LLR2bit}(\mathbf{L}(1,:) + \mathbf{R}(1,:))$
2: $\hat{\mathbf{x}} \leftarrow \text{LLR2bit}(\mathbf{L}(n+1,:) + \mathbf{R}(n+1,:))$
3: **switch** *stopID* **do**
4:   **case** 1       ▷ practical stopping criterion
5:     **if** $\hat{\mathbf{u}} \cdot \mathbf{G} = \hat{\mathbf{x}}$ **then**
6:       **return** true
7:     **end if**
8:   **case** 2       ▷ perfect knowledge-based stop. condition
9:     **if** $\hat{\mathbf{u}} = \mathbf{u}$ **then**
10:      **return** true
11:    **end if**
12:  **case** 3       ▷ CRC-aided stop. condition
13:    **if** $\hat{\mathbf{u}}$ satisfiesCRCcheck **then**
14:      **return** true
15:    **end if**
16: **end switch**
17: **return** false

For all simulations in this work, $10^6$ codewords per SNR point were simulated to get "stable" B(L)ER curves. The error-rate performance of this multi-trellis-based BP decoder while using the previously mentioned (*practical stopping condition*) is shown in Fig. 2a and 2b, respectively. The multi-trellis BP has a better error-rate performance when compared to the conventional BP decoder. It can be also depicted from Fig. 2a and 2b that the use of more different permutations of the factor graph enhances the error-rate performance, up to a certain point, when enhancement is no longer possible with increasing the number of factor graphs used. In Fig. 2a and 2b, it is also shown that the gain in performance is due to the use of different permutations of factor graphs not a result of the increased number of BP iterations. One reason for this performance improvement is due to the different structure of loops from one factor graph permutation to another, as seen from Fig. 1c. As shown, two loops are highlighted in the unfolded factor graph, whereas the nodes belonging to one loop in a graph permutation are spread among different loops in another graph permutation. Eventually, this means that one factor graph might be better for a specific noise realization than another. Besides, in the high SNR region (i.e., error floor region), the main cause of errors are the stopping sets. Therefore, shuffling the used factor graphs effectively leads to shuffling the variable nodes of the decoding graph, and, thus, resulting in totally different stopping sets [17].

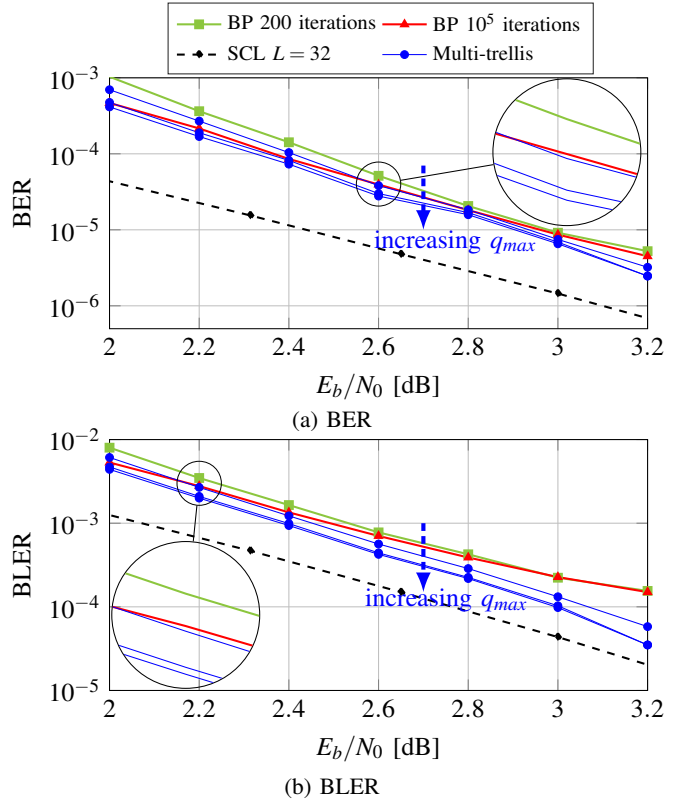

Fig. 2: BER and BLER comparison between BP, SCL with list size $L = 32$ and Multi-trellis BP for a $\mathscr{P}(2048,1024)$-code. Multi-trellis BP: the maximum number of iterations per trellis is $N_{it,max} = 200$, the maximum number of trellises is $q_{max} = \{2,5,10\}$, all with the practical stopping criterion.

## IV. EXTENSIONS AND COMPLEXITY

### A. Perfect knowledge-based termination

During our simulations, we noticed that in many cases in which the decoder fails, the decoder really converged to the set of correct information bits at a certain point but the stopping condition was not satisfied (i.e., $\hat{\mathbf{x}}$ and $\hat{\mathbf{u}}$ did not satisfy the stopping condition due to oscillating errors on both sides). This led to the conclusion that the performance of the multi-trellis BP decoder can be enhanced with a better stopping criterion, e.g., a genie to decide when to stop the iterations.

To test the validity of this lemma, the following experiment was conducted. The BP decoding stops when the estimated information bits in $\hat{\mathbf{u}}$ is equal to the transmitted information bits in $\mathbf{u}$, or when a maximum number of iterations per trellis $N_{it,max}$ over the maximum number of trellises $q_{max}$ is reached (second stopping condition in Algorithm 2). It is worth mentioning that this is not a practical decoder since it assumes perfect knowledge of the transmitted information bits. However, it provides a bound on the achievable decoding performance. The error-rate performance of this decoder (shown in Fig. 3a and 3b) approaches, and even outperforms, the SCL decoder performance (due to the perfect knowledge-based stopping criterion), indicating that one major cause of decoding failure in the multi-trellis BP decoder is the lack of proper stopping criterion. One can also see that the enhanced error-rate performance is due to the multiple different factor graph permutations used and not due to the increased number of BP iterations.

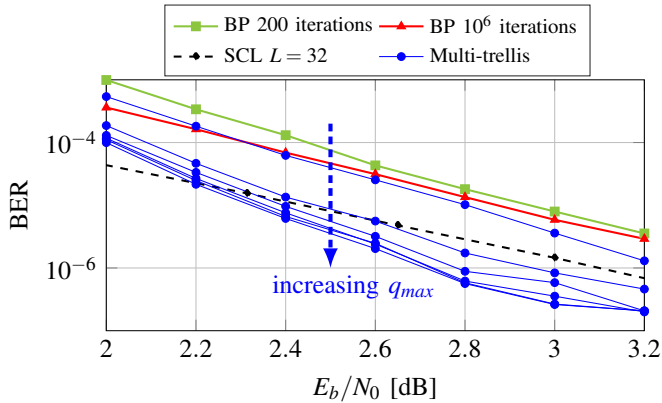

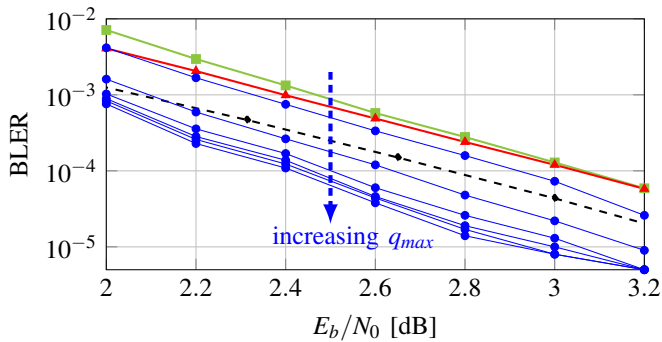

Fig. 3: BER and BLER comparison between BP, SCL $L = 32$ and Multi-trellis BP for a $\mathscr{P}(2048, 1024)$-code. Multi-trellis BP: $N_{it,max} = 200$, $q_{max} = \{2, 10, 100, 300, 500, 1000\}$, all iterative decoders use the perfect knowledge-based stopping criterion (i.e., the results provide a lower bound).

This means that for a specific noise realization, if the SCL decoder (which achieves the ML bound) can decode successfully, then there exists a polar factor graph such that the BP decoder can also decode successfully while using this specific factor graph with a carefully chosen stopping criterion.

### B. CRC-aided termination

Inspired by [3], one type of a practical genie that can be used is a high rate CRC code with $r$ redundancy bits. The CRC code can be considered as an outer code applied over the information bits, while the polar code is the inner code. The multi-trellis BP decoder iterations terminate if the CRC is satisfied, or when a maximum number of iterations per trellis $N_{it,max}$ over the maximum number of trellises $q_{max}$ is used (third stopping condition in Algorithm 2). Although we introduce a rate loss penalty $\left(\frac{r}{N}\right)$, this CRC-based stopping criterion is less complex than the previously mentioned practical stopping condition, which requires an additional polar re-encoding step. Fig. 4 shows that this CRC-aided decoding will help in approaching and even outperforming, in the high SNR region, the error-rate performance of the SCL decoder of a conventional polar code. Additionally, we provide results for BP decoding with the CRC-based stopping criterion for $N_{it,max} = 10^6$ iterations. This shows that the gain observed by the multi-trellis decoder is neither a result of the increased number of iterations nor of the CRC stopping condition. Thus the error-rate performance gain is due to the different permutations of the factor graphs used. It is worth mentioning that the concatenation of a polar code with a high rate CRC code has been proposed for the uplink control channel of the upcoming 5th generation mobile standard [2]. Thus, the proposed multi-trellis BP decoder is compatible with the (expected) standardized polar codes.

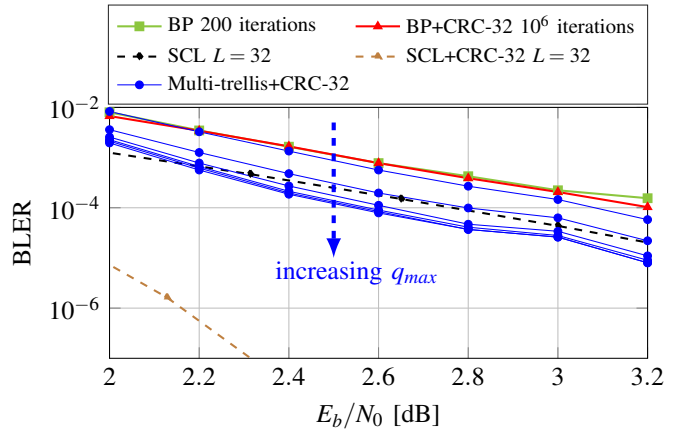

Fig. 4: BLER comparison between BP, BP+CRC, SCL $L = 32$, SCL+CRC $L = 32$ and Multi-trellis BP+CRC for a $\mathscr{P}(2048, 1024)$-code. The CRC used is $r = 32$ bits long. Multi-trellis BP: $N_{it,max} = 200$, $q_{max} = \{2, 10, 100, 300, 500, 1000\}$, with CRC-aided stopping criterion.

We want to emphasize that the performance of this CRC-aided decoder is not as good as the CRC-aided version of the SCL decoder (i.e., SCL-CRC). The CRC codes are well-suited for the task of picking the correct codeword from the list in the SCL decoder but in the multi-trellis BP decoder it is just used as a stopping guideline, thus, it is not adding much gain in the sense of information transfer (outer-inner iterative decoding strategy).

### C. Complexity

The main objective of this work is to propose this new variant of the BP decoder for polar codes and present its error-rate performance when compared to the conventional BP and SCL decoders. However, in its naive implementation, the price

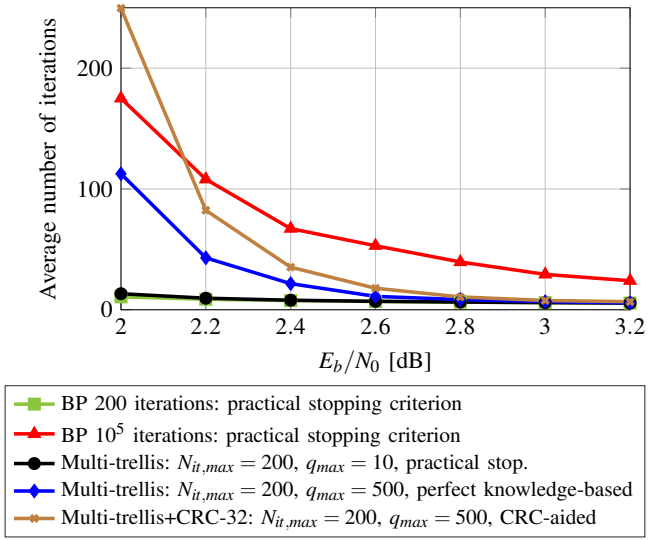

- ■— BP 200 iterations: practical stopping criterion
- ▲— BP $10^5$ iterations: practical stopping criterion
- ●— Multi-trellis: $N_{it,max} = 200$, $q_{max} = 10$, practical stop.
- ◆— Multi-trellis: $N_{it,max} = 200$, $q_{max} = 500$, perfect knowledge-based
- ✳— Multi-trellis+CRC-32: $N_{it,max} = 200$, $q_{max} = 500$, CRC-aided

Fig. 5: Average number of iterations performed per iterative decoding algorithm; polar code $\mathscr{P}(2048, 1024)$.

to pay is a rather high decoding complexity which we briefly evaluate next.

As can be seen from Fig. 5, the average number of performed decoding iterations does not drastically increase under multi-trellis decoding in the SNR range of interest, as the decoder typically operates in the low error-rate region. Obviously, the average number of iterations depends on the BLER of the plain BP decoder as each block failure causes additional iterations over the permuted trellises. Thus, decoding over multiple trellises is only required in some rare cases.

In the CRC-aided multi-trellis decoder, the complexity in the low SNR region is high, because the decoder will stop iterating when the CRC is satisfied, which is very unlikely due to bad channel conditions (i.e., the CRC-aided stopping condition is stricter than the practical stopping condition).

Various ideas can be implemented in order to reduce the complexity of the proposed decoder. The concept of a partitioned successive cancellation list (PSCL) decoder [20] can be used in this decoder where we treat a polar code of length $N$ as two polar codes of length $\frac{N}{2}$ (i.e., two partitions), and the proposed decoder is applied over each partition. This will reduce the complexity indeed, but on the expense of error-rate performance.

Furthermore, wasting useless iterations on non-converging factor graphs can be later avoided by using an LLR-based metric which quantifies the convergence behavior of a specific factor graph permutation. This will certainly reduce the complexity (and maximum latency) by a significant factor via skipping some (non-useful) permutations only after performing a few BP iterations over them.

## V. CONCLUSION

In this paper, we presented a new variant of the BP decoder for polar codes based on different permutations of the polar code factor graph. The proposed decoder, with a perfect knowledge-based termination criterion, approaches the error-rate performance of the state-of-the-art SCL decoder, suggesting that the current early BP decoding stopping criteria are not yet optimum and can be further optimized. A CRC-aided version of this decoder is proposed, which can outperform, in the high SNR region, the state-of-the-art SCL decoder of a plain polar code. Obviously, a CRC is a well-suited "genie" in the context of SCL decoding, and thus the performance of the CRC-aided SCL decoder is still better than any iterative decoding-based scheme for polar codes. Yet, to the best of our knowledge, the permuted factor graph-based BP decoder presented in this paper is the best iterative decoder for polar codes known thus far.